\documentclass[12pt,preprint]{aastex}
\usepackage{color}
\shorttitle{A NEW DYNAMICAL MASS OF THE VIRGO CLUSTER}
\shortauthors{Lee et al.}
\begin{document}
\title{A NEW DYNAMICAL MASS MEASUREMENT FOR THE VIRGO CLUSTER USING THE RADIAL VELOCITY PROFILE 
OF THE FILAMENT GALAXIES}
\author{Jounghun Lee\altaffilmark{1}, 
Suk Kim\altaffilmark{2}, and Soo-Chang Rey\altaffilmark{2}}
\altaffiltext{1}{Astronomy Program, Department of Physics and Astronomy, Seoul National University, 
Seoul 151-742, Korea
\email{ jounghun@astro.snu.ac.kr}}
\altaffiltext{2}{Department of Astronomy and Space Science, Chungnam National University,
Daejeon 305-764, Korea}
\begin{abstract} 
The radial velocities of the galaxies in the vicinity of a cluster shows deviation from the pure Hubble flow due to their 
gravitational interaction with the cluster. 
According to a recent study of Falco et al. (2014) based on a high-resolution N-body simulation, the radial velocity profile of the galaxies 
located at distances larger than three times the virial radius of a neighbor cluster can be well approximated by a universal formula 
and could be reconstructed from direct observables provided that the galaxies are distributed along one dimensional filament. They 
suggested an algorithm for the estimation of the dynamic mass of a cluster $M_{\rm v}$ by fitting the universal formula from the simulation 
to the reconstructed radial velocity profile of the filament galaxies around the cluster from observations. 
We apply the algorithm to two narrow straight filaments (called the Filament A and B) that were identified recently by Kim et al. (2015) 
in the vicinity of the Virgo cluster from the NASA-Sloan-Atlas catalog.  The dynamical mass of the Virgo cluster is estimated to be  
$M_{\rm v}=(0.84^{+2.75}_{-0.51})\times10^{15}\,h^{-1}M_{\odot}$ and $M_{\rm v}= (3.24^{+4.99}_{-1.31})\times 10^{15}\,h^{-1}M_{\odot}$ 
for the cases of the Filament A and B, respectively. We discuss observational and theoretical systematics intrinsic to the method of 
\citet{falco-etal14} as well as the physical implication of the final results.  
\end{abstract}
\section{INTRODUCTION}\label{sec:intro}

The simplest and probably the most reasonable explanation for the current acceleration of the universe is the anti-gravitational effect of 
the cosmological constant ($\Lambda$) that is supposed to be dominant at the present epoch. The fundamental assumption that underlies 
this explanation is that the general relativity (GR) is the universal law of gravity, the validity of which has been confirmed by recent 
cosmological observations \citep{reyes-etal10,wojtak-etal11,rapetti-etal13}. Besides, the latest news from the Planck CMB (Cosmic 
Microwave Background) mission has drawn the following bottom line \citep{planck_mg15}: 
The risk that one has to take by accepting the $\Lambda$CDM ($\Lambda$+cold dark matter) cosmology founded upon GR 
is lower than that by accepting any other alternative. 

Nevertheless, the conceptual difficulty to acquiesce in the extremely fine-tuned initial conditions of the universe associated 
with $\Lambda$ still dares the cosmologists to take a high risk of suggesting modified gravity (MG) models and developing independent 
tests of GR. One of the most plausible tests of GR on the cosmological scale is to compare the lensing mass of a cluster with its dynamic 
mass \citep[e.g., see][]{schmidt-etal09,schmidt10,zhao-etal11}.  The former is estimated from the weak gravitational lensing effect 
generated by the mass of a cluster on the shapes of the foreground galaxies, while the latter is conventionally derived from the velocity 
dispersions of the central satellites of a cluster. A substantial discrepancy between the two mass measurements for a cluster, 
if found and accurately measured, would challenge the key tenet of GR.

However, it is a formidable task to measure the dynamical mass of galaxy clusters with high accuracy. The success of the conventional 
method of using the satellite velocity dispersions is contingent upon the completion of relaxation in the dynamical state and the spherical 
symmetry in the shape. There is another method for the dynamical mass measurement of the clusters that uses the peculiar velocities 
of the neighbor galaxies located beyond the virial radius of a cluster. This method can be applied even to those clusters which have yet 
to be completely relaxed without having spherical symmetry \citep{corbett-etal14}. However, large uncertainties inherently involved in the 
measurements of the peculiar velocity field remain as a technical bottleneck that stymies the accurate measurement of the dynamic 
mass of a cluster from the peculiar velocities of the neighbor galaxies 
\citep[see also][]{corbett-etal14,gronke-etal14b,hellwing-etal14,zu-etal14}

\citet{falco-etal14} devised a novel method to measure the dynamic mass of a cluster by using the {\it universal} radial velocity profile of 
the filament galaxies around the cluster.  The sole condition for  the practical success of their method is that the neighbor galaxies 
should be distributed along a narrow filament.  Since it has been theoretically and observationally shown that the formation of galaxy 
clusters preferentially occur at the junction of the cosmic filaments \citep[e.g., see][]{web96,dietrich-etal12}, the majority of the galaxy 
clusters should meet this condition, being eligible for the application of this method.  
The method of \citet{falco-etal14} to measure the virial mass of a cluster from the radial velocity profile of the filament galaxies is applicable 
even to those clusters which have yet to complete its relaxation process. In this respect their method fits especially well for the Virgo cluster 
that still undergoes merging events. Furthermore, the measurements of the redshifts and equatorial coordinates of the galaxies in the 
vicinity of the Virgo cluster were done with high accuracy thanks to their proximity, which will help reduce uncertainties in the dynamical 
mass measurement with the method of \citet{falco-etal14} for the Virgo cluster. 
The goal of this paper is to test this new method against the Virgo cluster around which the narrow filamentary structures were recently 
identified by \citet{kim-etal15}. 

The upcoming sections will present the following. A review of the method of \citet{falco-etal14} in Section \ref{sec:f14}. 
A reconstruction of the radial velocity profile of the filament galaxies around the Virgo cluster and a new dynamical mass 
estimate for the Virgo cluster by comparing the reconstructed profile with the numerical formula from \citet{falco-etal14} in Section 
\ref{sec:mv}. Discussions on the final results and on the future works for the improvements in Section \ref{sec:con}. 
Throughout this paper, we assume a GR+$\Lambda$CDM cosmology with $\Omega_{m}=0.26,\ \Omega_{\Lambda}=0.74,\ h=0.73$ 
and zero curvature, to be consistent with \citet{falco-etal14}. 
 
\section{METHODOLOGY : A BRIEF REVIEW}
\label{sec:f14}

The galaxies located in the vicinity of a cluster must feel the gravity of the cluster even in the case that they are not 
the satellites bound to the cluster, and thus their motions relative to the cluster should deviate from the pure Hubble flows. 
To quantify the degree of the deviation of the galaxy motions from the Hubble flows and its dependence on the cluster 
mass, \citet{falco-etal14} determined the mean radial velocity profile of the galaxies in the vicinity of the clusters  
from a high-resolution N-body simulation for the GR+$\Lambda$CDM cosmology. Their numerical experiment has discovered that 
the mean radial velocity profile of the neighbor galaxies located in the zone where the effect of the Hubble expansion exceeds 
that of the gravitational attraction of a cluster is well approximated by the following universal formula:  
\begin{equation}
\label{eqn:vr}
v_{r}(d; z, M_{\rm v}) = H(z)\, d - 0.8\,V_{\rm v}\,\left(\frac{r_{\rm v}}{d}\right)^{n_{\rm v}}\, ,
\end{equation}
where $v_{r}$ is the mean radial velocity at (comoving) separation distance $d$ from the cluster center, $M_{\rm v}$, $r_{\rm v}$ and 
$V_{\rm v}$ are the virial mass, radius and velocity of the cluster, respectively, and $H(z)$ is the Hubble parameter at redshift $z$. 
The first term in the right-hand side represents a pure Hubble flow from the cluster center while the second term corresponds to 
the mean peculiar velocity induced by the gravitational attraction of the cluster. 

According to \citet{falco-etal14}, the radial velocity profile is universal in the respect that the power-law index, $n_{\rm v}$,  
in Equation (\ref{eqn:vr})  has a constant value of $0.42$, being independent of $M_{\rm v}$ and $z$. 
They also showed that the zone where the radial velocities of the galaxies are well approximated by Equation (\ref{eqn:vr}) 
with $n_{\rm v}=0.42$ corresponds to the range of $3\,r_{\rm v}\le d\le 8\,r_{\rm v}$.  Adopting the conventional definition that $r_{\rm v}$ 
is the radius of a sphere within which the mass density reaches $\Delta_{c}$ times the critical density of the Universe 
$\rho_{c}\equiv 3H^{2}/(8\pi\,G)$, \citet{falco-etal14} used the following simple relations to calculate the virial mass $M_{\rm v}$ and 
the virial velocity $V_{\rm v}$ for Equation (\ref{eqn:vr}): 
\begin{equation}
\label{eqn:mv_rv}
M_{\rm v} = \frac{4\pi}{3}\Delta\,\rho_{c}\,r^{3}_{\rm v},\ \quad V^{2}_{\rm v} = \frac{G\, M_{\rm v}}{r_{\rm v}}\, .
\end{equation}
The value of $\Delta_{c}$ was set at $93.7$ in accordance with the formula of 
$\Delta_{c}=18\pi^{2}+82[\Omega_{m}(z)-1]-39[\Omega_{m}(z)-1]^{2}$ given by \citet{BN98}.

\begin{figure}[b]
\begin{center}
\epsscale{1.0}
\plotone{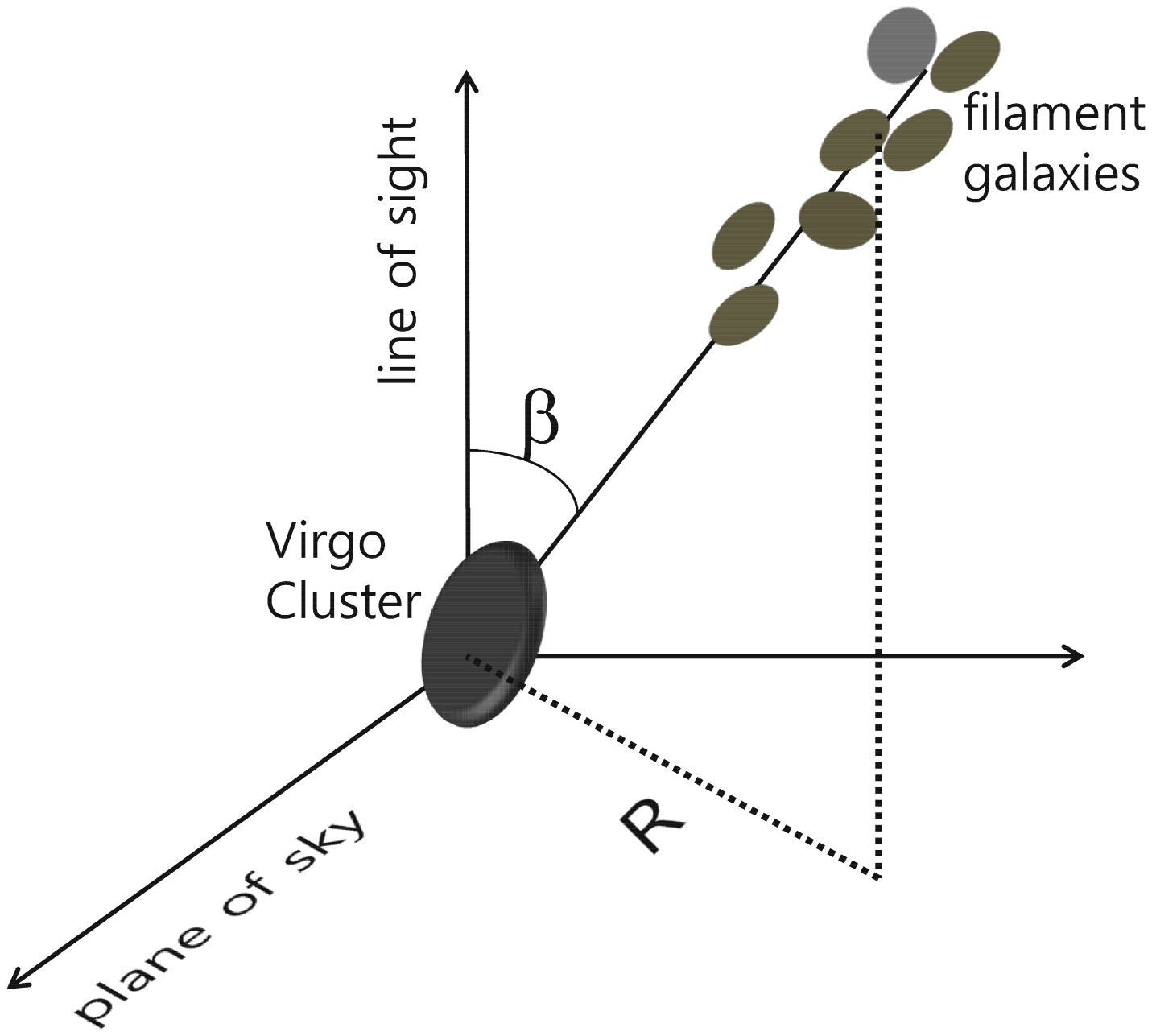}
\caption{Illustration of a three dimensional configuration of a filamentary structure inclined at an angle $\beta$ to 
the Virgo cluster. The projected distance $R$ of a filament galaxy from the Virgo center in the plane of sky perpendicular 
to the line of sight direction toward the Virgo can be measured from information on their equatorial coordinates.}
\label{fig:illustration}
\end{center}
\end{figure}

\citet{falco-etal14} suggested that the virial mass of a massive cluster $M_{v}$ be estimated by comparing the observed radial velocity 
profile of its neighbor galaxies to Equation (\ref{eqn:vr}). However, since the real values of $v_{r}$ and $d$ are not directly measurable, 
it is difficult to determine the radial velocity profiles of the galaxies from observational data. They claimed that this difficulty should be 
overcome by using the anisotropic spatial distribution of the galaxies belonging to a one-dimensional filament in the vicinity of a cluster.  
Three dimensional separation distance $d$ and the radial velocity $v_{r}$ of the filament galaxies can be estimated from 
two direct observables, the projected distance $R$ in the plane of the sky and the line-of-sight velocity $v_{\rm los}$, as 
$R=\sin\beta\, r$ and $v_{r}=v_{\rm los}/\cos\beta$, respectively, where $\beta$ is the angle at which the filament is inclined to the line of 
sight direction to a cluster (see Figure \ref{fig:illustration}). Finally, Equation (\ref{eqn:vr}) was rewritten in terms of the observables as  
\begin{equation}
\label{eqn:2dvr}
v_{\rm los}(R,\beta,M_{\rm v}) = 
\cos\beta\left[ H(z)\frac{R}{\sin\beta} - 0.8V_{\rm v}\left(\frac{R}{\sin\beta\,r_{\rm v}} \right)^{n_{\rm v}}\right]\, . 
\end{equation}

Since the galaxy clusters usually form at the junction of the cosmic filaments, this new scheme of \citet{falco-etal14} seems quite plausible. 
However, the validity of Equation (\ref{eqn:2dvr}) depends strongly on the geometrical shape of the filament. 
The more straight and narrow a filament is, the better Equation (\ref{eqn:vr}) is approximated by Equation (\ref{eqn:2dvr}).  
Figure \ref{fig:illustration} illustrates a three dimensional configuration of a filamentary structure around a cluster inclined at an angle 
$\beta$ to the line of sight direction to the cluster, defining the projected distance $R$ between a filament galaxy and the cluster center 
in the plane of sky.
\citet{falco-etal14} performed a numerical test of Equation (\ref{eqn:2dvr}) by applying it to three cluster-size halos around which filamentary 
structures were found. The estimated values of $M_{\rm v}$ turned out to agree very well with the true virial masses for all of the three 
cases, which numerically verified the validity of Equations (\ref{eqn:2dvr}).

\citet{falco-etal14} also performed an observational test of their algorithm by applying it to the sheet structures identified around the Coma 
cluster. 
Despite their claim that the dynamic mass of the Coma cluster determined by their method turned out to be consistent with the previous 
estimates, their method, strictly speaking, is valid only for the galaxies distributed along a one-dimensional filamentary 
structure that interacts gravitationally with the cluster. In fact, it is obvious from Equation (\ref{eqn:2dvr}) that large uncertainties should 
contaminate the measurement of the cluster mass if the radial velocity profile is reconstructed from the sheet galaxies.  

\section{ESTIMATING THE DYNAMIC MASS OF THE VIRGO CLUSTER}
\label{sec:mv}

\subsection{The Zero-th Order Approximation}\label{sec:mv0}

\citet{kim-etal15} developed an efficient technique to identify a filamentary structure whose galaxies are under the 
gravitational influence of a neighbor cluster. The merit of their filament-finding technique is that it requires information only on 
observable redshifts and two dimensional positions of the galaxies projected onto the plane of sky. Applying this filament-finding 
technique to  the NASA-Sloan-Atlas catalog of the local galaxies\footnote{given in http://www.nsatlas.org/data}, 
\citet{kim-etal15} have identified two narrow filamentary structures to the east and to the west of the Virgo cluster (say, the Filament A and 
B, respectively). Both of the Filament A and B were found to consist mainly of the dwarf galaxies with absolute $r$-band magnitudes 
$M_{r}> -20$.   For a detailed description about how the Filament A and B were identified from the NASA-Sloan-Atlas catalog, see \citet{kim-etal15}.

\begin{figure}
\begin{center}
\epsscale{1.0}
\plotone{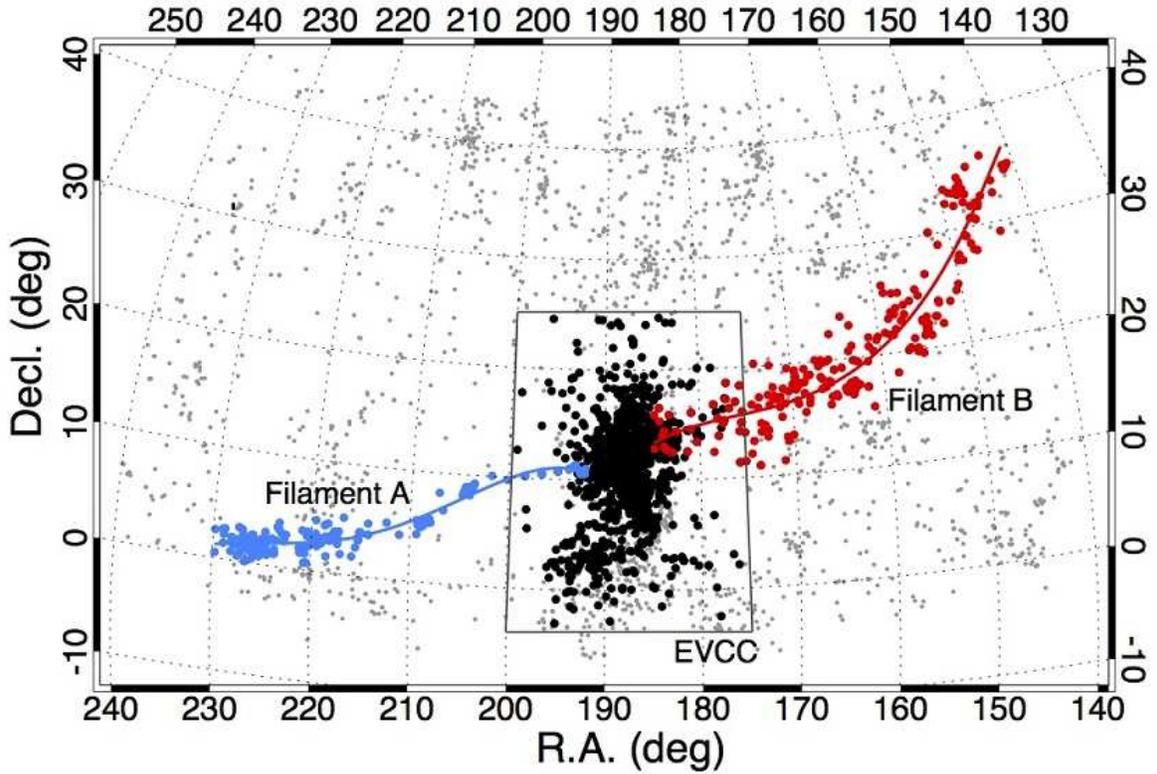}
\caption{Spatial distribution of the Filament A and B around the Virgo cluster in the equatorial 
coordinate system. The black dots correspond to the galaxies of the Virgo Cluster while the filled blue (red) dots 
correspond to the galaxies belonging to the Filament A (B). The grey dots represent the galaxies that are not associated 
with the Virgo cluster nor filaments.The fitted lines of the galaxies in the Filament A and B are also denoted by the solid curves.}
\label{fig:dec_ra}
\end{center}
\end{figure}

Figure \ref{fig:dec_ra} plots the right asensions (R.A.) and  declinations (decl.) of the galaxies belonging to the filament 
A and B, as filled blue and red dots, respectively. The certain and possible member galaxies  of the Virgo cluster from the 
Extended Virgo Cluster Catalog (EVCC) \citep{kim-etal14} are shown as the black and gray dots, respectively, inside an open 
rectangular box. The gray dots outside the box represent the field galaxies that do not belong to the EVCC 
nor to the two filaments but in the redshift range of $500\le cz/[{\rm km}\,s^{-1}] \le 2500$. 
Note  that the Filament A appears to be more straight than the Filament B in the equatorial plane. 

\begin{figure}
\begin{center}
\epsscale{1.0}
\plotone{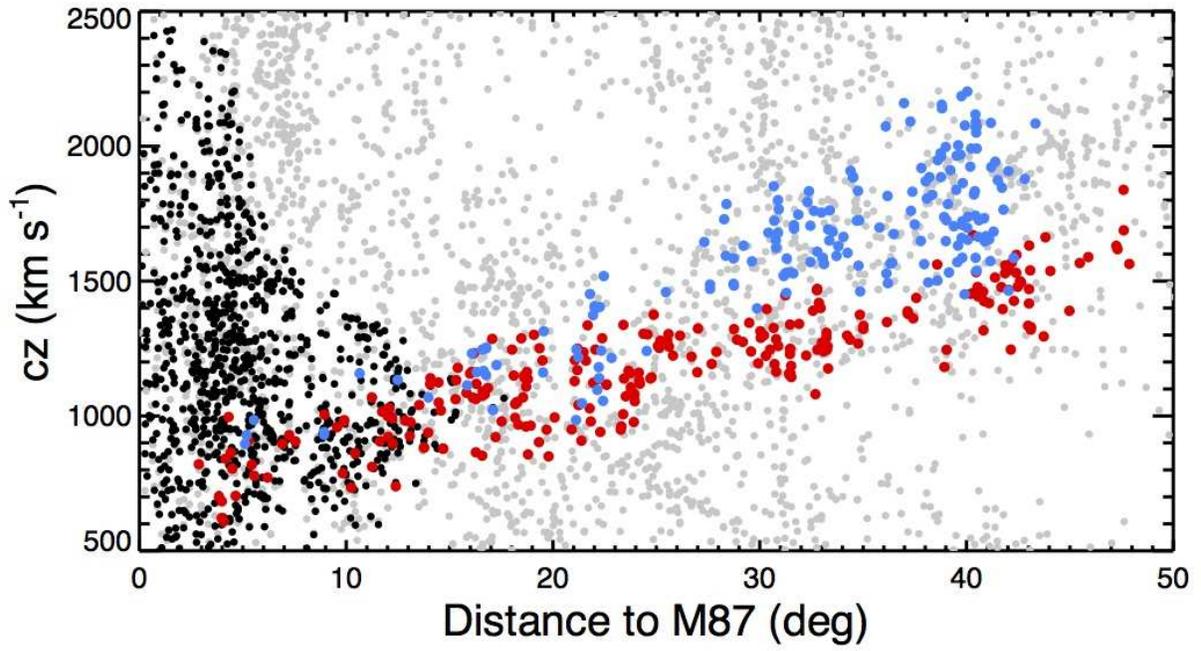}
\caption{Distances of the Filament galaxies from the center of the Virgo cluster versus their recession velocities. 
The symbols are the same as in Figure \ref{fig:dec_ra}.}
\label{fig:cz_dis}
\end{center}
\end{figure}

Showing that the mean radial velocities of the galaxies belonging to each filament are similar to the system velocity of the Virgo cluster, 
\citet{kim-etal15} have confirmed that the filament galaxies are in gravitational interaction with the Virgo cluster. 
Figure \ref{fig:cz_dis} plots the redshifts of the galaxies belonging to the Filament A and B as a function of their projected distances 
in unit of degree from the M87 (i.e., the Virgo center) as filled blue and red dots, respectively. As can be seen, the redshift variation 
of the filament galaxies $\Delta\,cz$ at the constant $R$ are less than $1000$ km/s.  

Denoting the equatorial positions (i.e., R.A. and decl.) of a filament galaxy and the Virgo center as $(\alpha_{\rm g},\ \delta_{\rm g})$ 
and $(\alpha_{\rm cl},\ \delta_{\rm cl})$, respectively, we determine the projected distance $R$ between 
each filament galaxy from the Virgo center in the plane of sky (see Figure \ref{fig:illustration}) as 
$R = c\,z_{\rm cl}\left(\cos\theta_{\rm g}\cos\theta_{\rm cl} + \sin\theta_{\rm g}\sin\theta_{\rm cl}\cos\Delta\phi_{\rm g}\right)$ where 
$\theta_{\rm g}=\pi/2-\delta_{\rm g}$, $\theta_{\rm cl}=\pi/2-\delta_{\rm cl}$, $\Delta\phi_{\rm g}=\alpha_{\rm g}-\alpha_{\rm cl}$, and 
$z_{\rm g}$ and $z_{\rm cl}$ are the redshifts of the filament galaxy and the Virgo center, respectively.  
The line-of-sight velocity magnitude $v^{\rm ob}_{\rm los}$ of each filament galaxy at a projected distance, $R$, from the Virgo center  
is approximated as $c\vert\,z_{\rm g}(R) - z_{\rm cl}\,\vert$.

Equation (\ref{eqn:2dvr}) is valid only for those filament galaxies whose separation distances from the cluster center are larger than three 
times the virial radius of the Virgo cluster according to \citet{falco-etal14}.  Taking the typical cluster size $2\,h^{-1}$Mpc as a conservative 
upper limit on the virial radius of the Virgo cluster, we select only those filament galaxies which satisfy the condition of $R\ge 6\,h^{-1}$Mpc. 
A total of $130$ and $96$ galaxies are selected in the Filament A and B, respectively, to each of which the method of \citet{falco-etal14} 
is applied as follows. 

Setting the power-law index $n_{\rm v}$ at the universal value of $0.42$, we compare the observed values of $v^{\rm ob}_{\rm los}$ of the 
filament galaxies with Equation (\ref{eqn:2dvr}). Assuming a flat prior, we search for the best-fit values of $M_{\rm v}$ and $\beta$ which 
maximize the following posterior density function $p(M_{\rm v}, \beta | {\rm data})$: 
\begin{equation}
\label{eqn:pos_mv_beta}
p(M_{\rm v}, \beta | v^{\rm o}_{\rm los},R_{i}) \propto \exp\Bigg{\{}-\sum_{i=1}^{N_{g}}
\frac{[v^{\rm o}_{\rm los}(R_{i}) - v_{\rm los}(R_{i}; M_{\rm v},\ \beta)]^{2}}{2\sigma^{2}_{\rm v}(N_{\rm g}-2)}\Bigg{\}}\, ,
\end{equation}
where $N_{\rm g}$ is the number of the selected filament galaxies,  $R_{i}$ is the projected distance of the $i$-th filament galaxy,  
$\sigma_{\rm v}$ represents errors involved in the measurement of $v^{\rm ob}_{los}(R_{i})$ from the redshift space. 
Given that $\sigma_{\rm v}$ include all systematic errors that must be existent but unknown, we set $\sigma_{\rm v}$ at unity for all of the 
selected filament galaxies. Here, the posterior probability density function $p(M_{\rm v}, \beta | v^{\rm o}_{\rm los},R_{i})$ is normalized to 
satisfy the condition of $\int dM_{\rm v}\int d\beta\, p(M_{\rm v}, \beta | v^{\rm ob}_{\rm los},\ R_{i}) = 1$. 

\begin{figure}[b]
\begin{center}
\epsscale{1.0}
\plotone{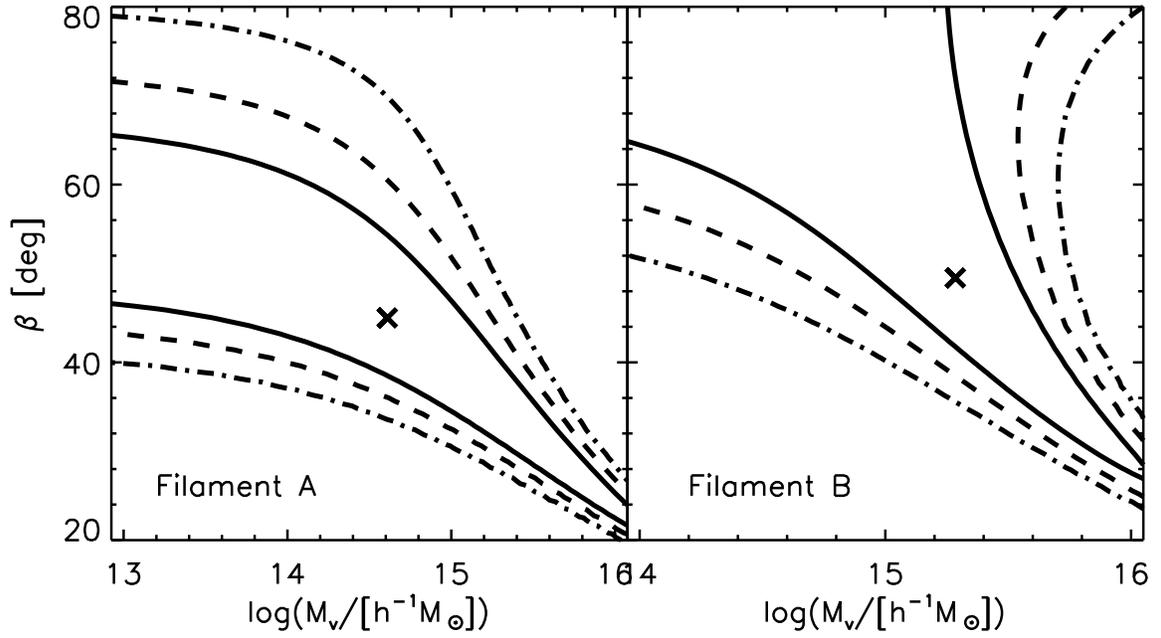}
\caption{$68\%$ (solid), $95\%$ (dashed) and $99\%$ (dot-dashed) confidence level contours of $\log M_{\rm v}$ and 
$\beta$ for a $\Lambda$CDM cosmology by comparing Equation (\ref{eqn:2dvr}) to the observed radial velocity profile 
of the galaxies in the Filament A and B in the left and right panels, respectively. In each panel the cross marks the 
best-fit $(\log M_{\rm v},\beta)$.}
\label{fig:mv_beta}
\end{center}
\end{figure}

The $68\%,\ 95\%$ and $99\%$ confidence level contours of $\log M_{\rm v}$ and $\beta$ obtained from the above fitting procedure 
for the case of the Filament A (B) are shown as solid, dashed and dot-dashed lines, respectively, in the left (right) panel of Figure 
\ref{fig:mv_beta}, which reveals the existence of a strong degeneracy between $M_{v}$ and $\beta$. To break this degeneracy, we would 
like to put some prior on the range of $\beta$ by making the zeroth-order approximation to the three dimensional positions of the 
filament galaxies relative to the Virgo cluster.  

In the zero-th order approximation of $d\approx d^{0}=c\,z_{\rm g}$ under the assumption of no peculiar velocity,  the three dimensional 
separation vector of each selected filament galaxy, ${\bf d}^{0}=(d_{1}^{0},\ d_{2}^{0},\ d_{3}^{0})$, can be expressed as 
$d_{1}^{0} = c\,z_{g}\sin\delta_{\rm g}\cos\alpha_{\rm g} - c\,z_{\rm cl}\sin\delta_{\rm cl}\cos\alpha_{\rm cl},\ 
d_{2}^{0} = c\,z_{g}\sin\delta_{\rm g}\sin\alpha_{\rm g} - c\,z_{\rm cl}\sin\delta_{\rm cl}\sin\alpha_{\rm cl},\ 
d_{3}^{0} = c\,z_{g}\cos\delta_{\rm g} - c\,z_{\rm cl}\cos\delta_{\rm cl}$. The zeroth-order approximation to the angle, $\beta$, 
between ${\bf d}$ and the line of sight direction to the Virgo cluster, say $\hat{\bf h}$, is now calculated as, 
$\beta\approx \beta^{0}\equiv \cos^{-1}\langle\hat{\bf d}^{0}\cdot\hat{\bf h}\rangle$ with 
$\hat{\bf d}^{0} \equiv {\bf d}^{0}/\vert{\bf d}^{0}\vert$ where the ensemble average is taken over the selected filament galaxies.   

The Filament A and B yield $\beta^{0}=36.8^{\circ}\pm 3.9^{\circ}$and $\beta^{0}=36.9^{\circ}\pm 5.6^{\circ}$, respectively. 
It is interesting to note that this result is in line with the previous observational evidences for the alignment tendency between 
the major axis of the Virgo cluster and the line-of-sight direction \citep{mei-etal07}. It is also consistent with the recent detection 
that the Virgo satellites tend to fall into the Virgo along the lie of sight directions aligned with the minor principal axis of the local 
velocity shear field \citep{lee-etal14}. 
Taking these zero-th order values as a prior,  we break the degeneracy between $M_{\rm v}$ and $\beta$ 
shown in Figure \ref{fig:mv_beta} and finally determine the dynamical mass of the Virgo cluster: 
$M_{\rm v}\approx (0.84^{+2.75}_{-0.51})\times 10^{15}\,h^{-1}M_{\odot}$ for the case of the Filament A and 
$M_{\rm v}\approx (3.2^{+5.0}_{-1.3})\times 10^{15}\,h^{-1}M_{\odot}$ for the case of the Filament B. 

\subsection{Radial Motions of the Filament Galaxies around the Virgo Cluster}\label{sec:vr}

It should be worth comparing our result with the previous measurements of the Virgo mass based on various methods.
For instance, \citet{urban-etal11} converted the X-ray temperature of the hot gas in the Virgo cluster to its mass via the mass-temperature 
scaling relation obtained by \citet{arnaud-etal05} and found $M_{200}/M_{\odot}\approx 1.4\times 10^{14}$. Here $M_{200}$ represents the 
mass enclosed by a spherical radius $r_{200}$ at which the mass density becomes $\rho(r_{200})=200\,\rho_{crit}$ where 
$\rho_{crit}\equiv 3H^{2}/8\pi G$.   
\citet{karachentsev-etal14} studied the infall motions of seven nearby galaxies located along the line-of-sight direction to the 
Virgo cluster whose distances were measured by means of the Tip of the Red Giant Branch (TRGB) method \citep{baade44,SC97} and 
found $M_{\rm 200}/M_{\odot}=(7.0\pm 0.4)\times 10^{14}$.  It was also claimed that the lensing mass of the Virgo cluster should be in the 
range of $\log\left(M_{200}/M_{\odot}\right)=13.87^{+0.42}_{-1.95}$ by employing the technique developed by \citet{kubo-etal09} to estimate 
the lensing mass of the galaxy clusters in the local universe   
(D. Nelson 2014, private communication\footnote{see also the presentation given in 
https://www.cfa.harvard.edu/~dnelson/doc/dnelson.fermilab.talk.pdf. }).

Noting that the definition of $M_{200}$ is different from that of the virial mass $M_{\rm v}$ considered here, we convert the values of 
$M_{200}/M_{\odot}$ reported in the previous works into $M_{\rm v}/(h^{-1}\,M_{\odot})$ for a fair comparison under the assumption that 
the mass density profile of the Virgo cluster is well approximated by the NFW formula \citep{nfw97} and list them in Table \ref{tab:mv}. 
As can be read,  our best-fit value $M_{\rm v}$ obtained from the Filament A is consistent with the previous dynamic mass estimated 
by the TRGB-distance method but substantially higher than the other two estimates. Meanwhile, for the case of the Filament B, 
the best-fit $M_{\rm v}$ is significantly off not only from the previous estimates but from the best-fit value obtained for the case of the 
Filament A.  Given that the method of \citet{falco-etal14} is strictly valid only for the galaxies distributed along thin straight filament, 
we suspect that  the bent shape of the Filament B should be responsible for the significant discrepancy on $M_{\rm v}$ 
between the two filaments and suggest that the Filament B should not be eligible for the application of the method of \citet{falco-etal14}.

Before excluding the Filament B from our analysis, it may be worth investigating how different the reconstructed velocity profile 
of the Virgo cluster is from the expected profile, i.e., the universal formula given by \citet{falco-etal14} when the mass of the Virgo 
cluster is fixed at some constant value $M_{\rm v}$. 
Treating $n_{\rm v}$ in Equation (\ref{eqn:2dvr}) as a free parameter, we search for the best-fit values of $n_{\rm v}$ and $\beta$ that 
maximize the following posterior distribution: 
\begin{eqnarray}
p(n_{\rm v}, \beta | v^{\rm ob}_{\rm los}, R_{i}) &\propto& p({\rm data} | n_{\rm v}, \beta) p(\beta) \, \nonumber \\
\label{eqn:pos_nv_beta}
&\propto& 
\exp\left\{-\frac{[v^{\rm o}_{\rm los}(R_{i}) - v_{\rm los}(R_{i}; n_{\rm v}, \beta)]^{2}}{2(N_{g}-2)}\right\}
\frac{1}{\sigma_{\beta^{0}}}\exp\left[-\frac{\left(\beta-\bar{\beta^{0}}\right)^{2}}{2\sigma^{2}_{\beta^{0}}}\right]\, .
\end{eqnarray}
Here, instead of setting $\beta$ at the fixed value $\bar{\beta^{0}}$, we assume a Gaussian prior on $\beta$, setting its 
mean and standard deviation at $\bar{\beta^{0}}$ and $\sigma_{\beta 0}$, respectively, both of which are determined in Section \ref{sec:mv0}.

\begin{figure}[ht]
\begin{center}
\epsscale{1.0}
\plotone{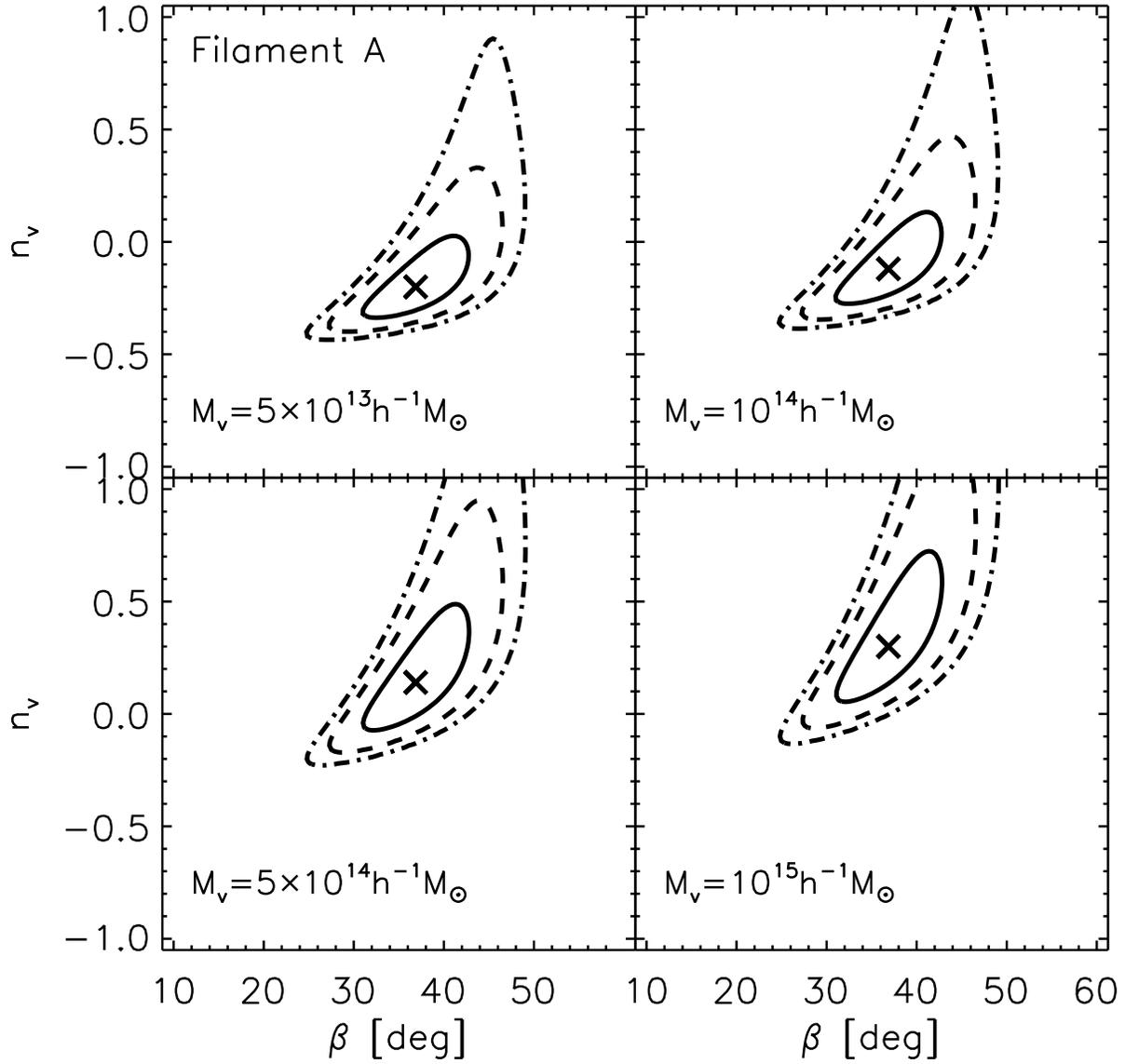}
\caption{$68\%$ (solid), $95\%$ (dashed) and $99\%$ (dot-dashed) confidence level contours of $n_{\rm v}$ and $\beta$ 
from the observed radial velocity profiles of the galaxies in the Virgo Filament A. The cross marks the best-fit 
$(n_{\rm v}, \beta$).}
\label{fig:nv_betaA}
\end{center}
\end{figure}

\begin{figure}[ht]
\begin{center}
\epsscale{1.0}
\plotone{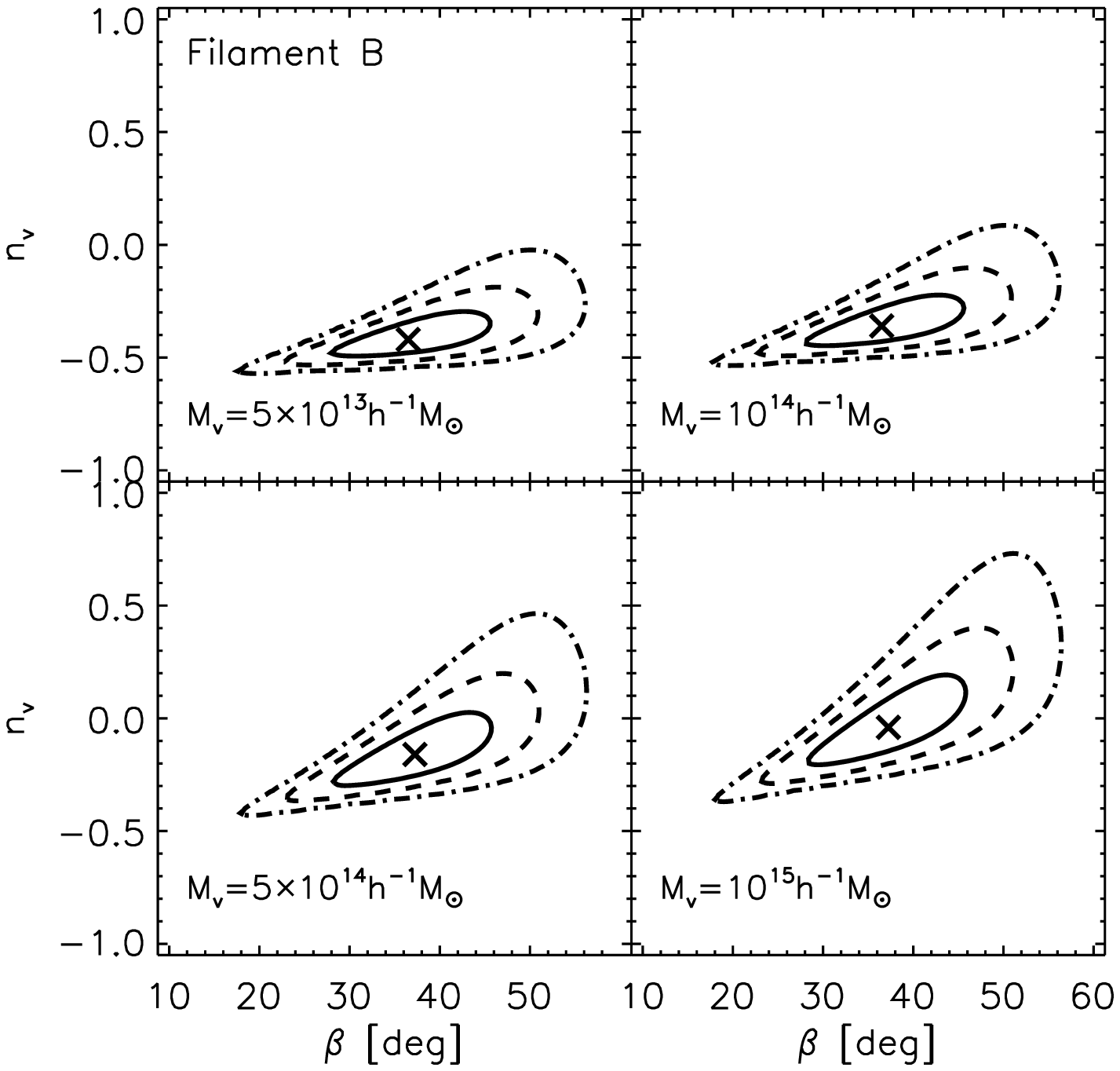}
\caption{ Same as Figure \ref{fig:nv_betaA} but for the case of the Filament B.}
\label{fig:nv_betaB}
\end{center}
\end{figure}

Figure \ref{fig:nv_betaA} shows  the $68\%,\ 95\%$ and $99\%$ confidence level contours of $\beta$ and $n_{\rm v}$ obtained 
from the Filament A as solid, dashed and dot-dashed lines, respectively, for the four different cases of the dynamical mass 
of the Virgo cluster: $M_{\rm v} = 5\times 10^{13},\ 10^{14},\ 5\times 10^{14},\ 10^{15}\, h^{-1}M_{\odot}$. 
As can be seen, for the cases of $M_{\rm v}< 10^{14}\,h^{-1}M_{\odot}$ that is consistent with the previous estimate of 
the lensing mass (see Table 1), the original value of $n_{\rm v}=0.42$ is located beyond the $95\%$ confidence level contours. 
For the case of $M_{\rm v}\ge 5\times 10^{14}\,h^{-1}M_{\odot}$ that is consistent with the dynamical mass estimate obtained by the 
TRGB method, $n_{\rm v}=0.42$ is included within the $68\%$ confidence level contour. Hence, one can conclude that the numerical 
formula of the radial velocity profile suggested by \citet{falco-etal14} matches fairly well with the reconstructed profile from observation 
for the case of the Filament A around the Virgo cluster, yielding the best-fit value of $M_{\rm v}$ consistent with the previous 
dynamical mass measurement for the Virgo. Figure \ref{fig:nv_betaB} plots the same as Figure \ref{fig:nv_betaA} but from the Filament B.  
As can be seen, for all of the four cases of $M_{\rm v}$,  the original value of $0.42$ stays outside the $95\%$ contours, indicating that 
Equation (\ref{eqn:2dvr}) fails to describe the radial motions of the galaxies in the Filament B. 

\begin{figure}[ht]
\begin{center}
\epsscale{1.0}
\plotone{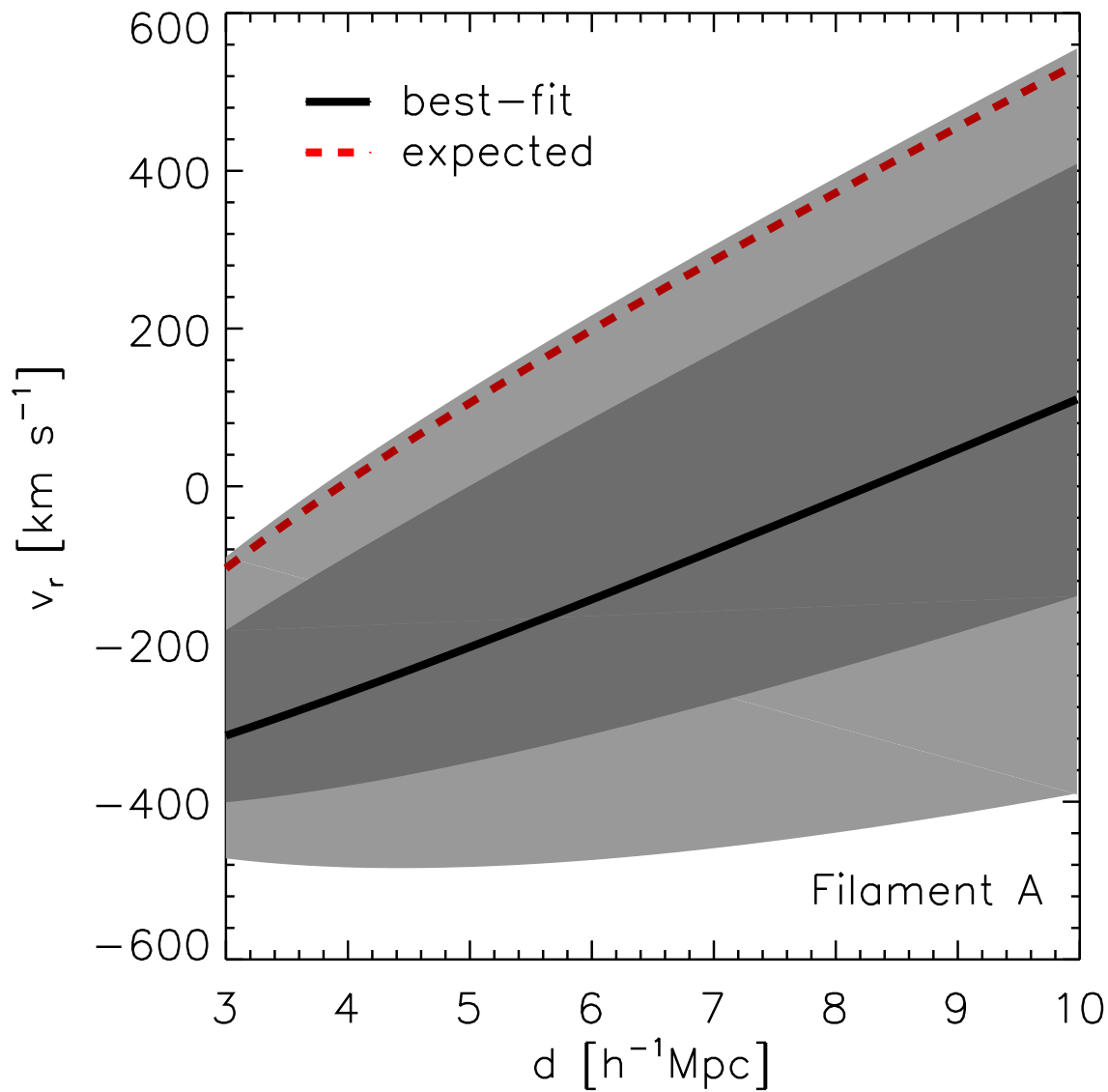}
\caption{Velocity profile of the Filament A galaxies around the Virgo. The black solid line corresponds to the best-fit velocity 
profile inferred from the observation data while the red solid line is the expected velocity profile of a galaxy cluster with 
mass of $10^{14}\,h^{-1}M_{\odot}$ in the $\Lambda$CDM cosmology. The regions in dark (light) gray colors 
represent the $1\sigma$ (2$\sigma$) scatters in the determination of the best-fit velocity profile} 
\label{fig:vprofileA}
\end{center}
\end{figure}

\begin{figure}[ht]
\begin{center}
\epsscale{1.0}
\plotone{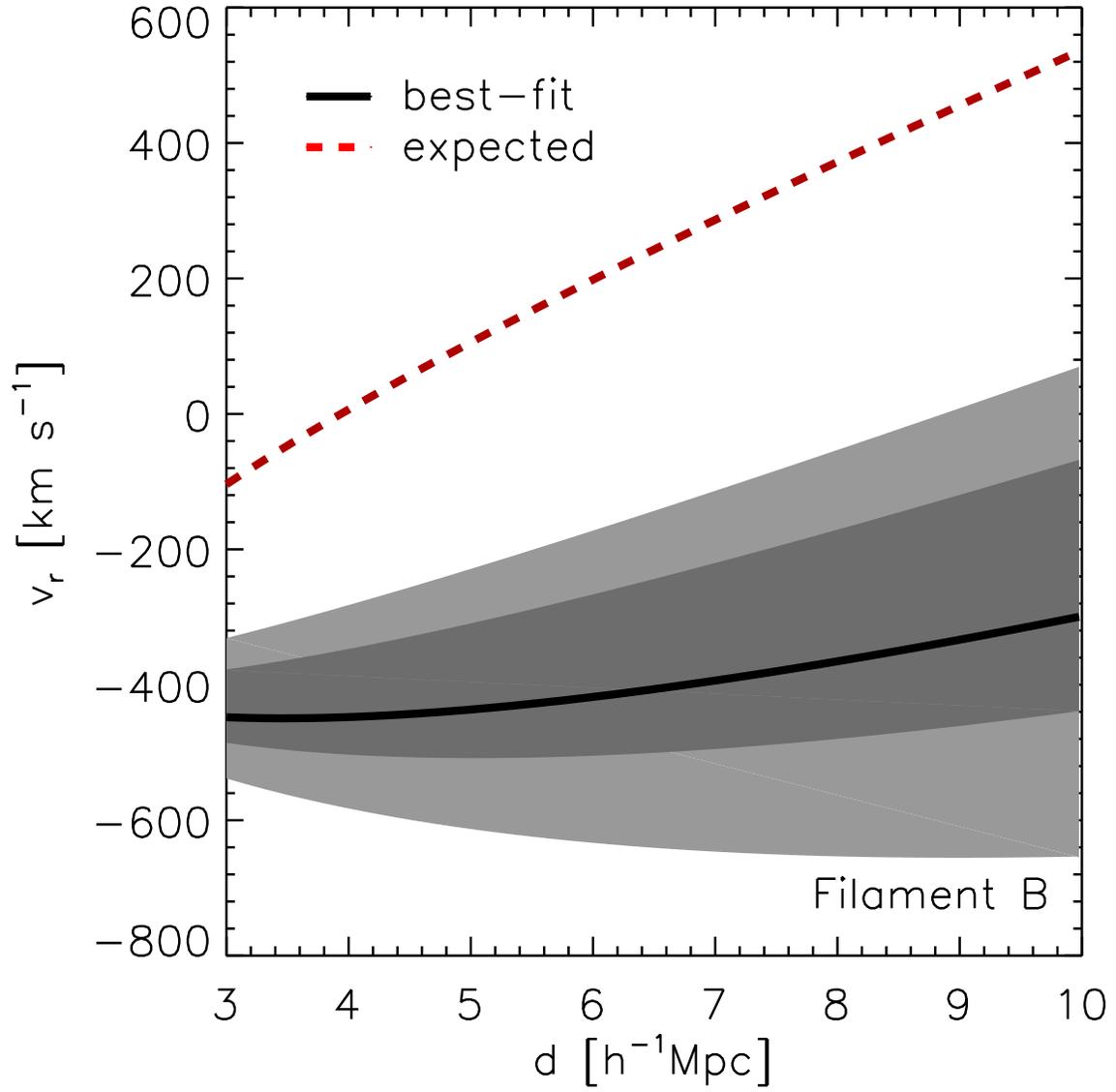}
\caption{Same as Figure \ref{fig:vprofileA} but for the case of the Filament B galaxies.}
\label{fig:vprofileB}
\end{center}
\end{figure}

We also show the radial velocity profile, $v_{r}(d)$,  reconstructed from the Filament A as black solid line in Figure \ref{fig:vprofileA}.  
To plot $v_{r}$, the Virgo mass $M_{\rm v}$ is set at $10^{14}\,h^{-1}M_{\odot}$ (close to the lensing mass), 
the slope $n_{\rm v}$ and the angle $\beta$ are set at the best-fit values determined by the above fitting procedure. 
The associated $1\sigma$ and $2\sigma$ uncertainties around the reconstructed profile are also shown as dark and light grey regions, 
respectively, and the red dashed line represents the expected profile with the fixed value of $n_{v}=0.42$ claimed by \citet{falco-etal14}. 
As can be seen, the expected profile is only marginally consistent with the reconstructed one if $M_{\rm v}=10^{14}\,h^{-1}M_{\odot}$. 
Figure \ref{fig:vprofileB} plots the same but for the case of the Filament B. It is clear that the difference between the numerical profile 
and the reconstructed one becomes much larger for the case of the Filament B. 

\section{DISCUSSION AND CONCLUSION}
\label{sec:con}

We have measured the dynamic mass of the Virgo cluster, $M_{\rm v}$, by applying the method of \citet{falco-etal14} to two narrow 
filaments (the Filament A and B) consisting mainly of dwarf galaxies located to the east and to the west of the Virgo cluster in the 
equatorial reference frame, which were recently identified by \citet{kim-etal15} from the NASA-Sloan-Atlas catalog. 
The method of \citet{falco-etal14} that can be applied even to the unrelaxed clusters like the Virgo is based on the numerical findings 
that the radial velocity profile of the galaxies at distances larger than three times the virial radius of a neighbor cluster has a universal 
shape and that it can be readily reconstructed from direct observables as far as the galaxies are distributed along a filamentary structure. 

To break a strong  degeneracy found between the value of $M_{\rm v}$ and the angle $\beta$, we have constrained the ranges of $\beta$ by 
making a zero-th order approximation in which the redshift-space positions of the filament galaxies are assumed to equal their real-space ones. 
Our estimates based on the method of \citet{falco-etal14} has yielded 
$M_{\rm v}=(0.84^{+2.75}_{-0.51})\times10^{15}\,h^{-1}M_{\odot}$ and $M_{\rm v}= (3.2^{+5.0}_{-1.3})\times 10^{15}\,h^{-1}M_{\odot}$ 
from the Filament A and B, respectively. 

Given that the Filament B appears not so straight as the Filament A in the equatorial plane, the substantial difference between the two 
results on the value of $M_{\rm v}$ from the Filament A and B has been suspected as an indication that the Filament B is not eligible 
for the algorithm of \citet{falco-etal14} whose success depends sensitively on the degree of the filament straightness. 
The method of \citet{falco-etal14} itself also suffers from a couple of unjustified assumptions. First of all, it assumes the filaments to be 
extremely narrow and straight, which is an obvious over-simplication of the true shapes of the filaments. Second of all, in their original 
study, the universal radial velocity profile was obtained by taking an average over all the particles and halos around the clusters but not only 
over the filament halos. If only a subsample composed of the filament galaxies is used to reconstruct the radial velocity, then it would not 
necessarily agree with the average universal profile. Besides, the zero-th order approximation that we have made for $\beta$ in our 
estimation of the Virgo mass with the method of \citet{falco-etal14} is an additional simplification of the reality. If a broader range of 
$\beta$ were considered, then the mismatches between the two masses would be likely reduced.

Our estimate of $M_{\rm v}$ from the Filament A has turned out to be consistent with the previous dynamical mass estimates based on 
the TRGB-distance method by \citet{karachentsev-etal14} but three to four times higher than the other mass estimates based on the weak 
gravitational lensing effect (D. Nelson 2014 in private communication) and the X-ray temperature of hot gas \citep{urban-etal11}.  
The usual suspect for this disagreement between the mass estimates  for the Virgo cluster is the contamination of the measurements due to the 
presence of systematics associated with such over-simplified assumptions as the spherical NFW density profile \citep{nfw97} in the lensing mass 
estimates, the hydrostatic equilibrium condition for the X-ray temperature measurement, and etc.

Another more speculative explanation for the difference between the lensing and the dynamic mass measurements for the Virgo cluster 
is the failure of GR. We have shown that if the dynamic mass of the Virgo cluster is assumed to be identical to the lensing mass, then the 
reconstructed radial velocity profile from observational data appears to less rapidly decrease with the distance than the universal radial 
profile from the simulation for the GR+$\Lambda$CDM cosmology. The less rapid decrease of the radial velocity profile with distance 
indicates higher peculiar velocities of the filament galaxies at large distances than predicted for the case of GR.
In fact, several numerical works already explored how much the peculiar velocities of galaxies around massive clusters would be enhanced 
in the presence of modified gravity (MG) \citep{lam-etal12,corbett-etal14,gronke-etal14b,zu-etal14}. 
Before exploring this speculative idea, however, it will be necessary to study comprehensively how the systematics produced by the 
over-simplified assumptions mentioned in the above would affect the reconstruction of the radial velocity profile of the filament galaxies 
in the vicinity of the clusters.  Our future work is in this direction. 

\acknowledgments

JL thanks D.Nelson for very helpful discussion.
This work was supported by the research grant from the National Research Foundation of Korea to the Center for 
Galaxy Evolution Research  (NO. 2010-0027910).  JL also acknowledges the financial support by the Basic Science Research 
Program through the National Research Foundation of Korea (NRF) funded by the Ministry of Education (NO. 2013004372).
S.C.R acknowledges the support by the Basic Science Research Program through the National Research Foundation of
Korea (NRF) funded by the Ministry of Education, Science, and Technology (NRF-2012R1A1B4003097).
S.K. acknowledges support from the National Junior Research Fellowship of NRF (No. 2011-0012618).

\clearpage
\clearpage
\clearpage
\clearpage
\clearpage
\clearpage
\clearpage
\clearpage
\clearpage
\begin{deluxetable}{cccc}
\tablewidth{0pt}
\setlength{\tabcolsep}{5mm}
\tablecaption{Observable, virial mass estimate of the Virgo cluster in unit of $10^{14}\,h^{-1}M_{\odot}$ with $h=0.73$ 
and literature that reports the estimate.}
\tablehead{indicator & $M_{\rm v}$ &  literature  \\
 & ($10^{14}\,h^{-1}M_{\odot}$) &  }
\startdata
X-ray temperature  & $ 1.50 $ & Urban et al. (2011)\\
weak lensing effect  & $0.81^{+1.29}_{-0.80} $ & Nelson (2014)\tablenotemark{a} \\
TRGB distance & $7.73\pm 0.43$ & Karachentsev et al. (2014) 
\enddata
\tablenotetext{a}{https://www.cfa.harvard.edu/~dnelson/doc/dnelson.fermilab.talk.pdf} 
\label{tab:mv}
\end{deluxetable}
\end{document}